\documentclass[11pt]{article}
\usepackage[T1]{fontenc}
\usepackage{lmodern}
\usepackage{amsmath,amssymb,amsthm,mathtools}
\usepackage[margin=1in]{geometry}
\usepackage{booktabs}
\usepackage{float}
\usepackage{array}
\usepackage{tikz}
\usetikzlibrary{shapes.geometric,decorations.pathreplacing}
\usepackage{microtype}
\usepackage[hidelinks]{hyperref}
\newtheorem{theorem}{Theorem}
\newtheorem{lemma}[theorem]{Lemma}
\newtheorem{proposition}[theorem]{Proposition}
\newtheorem{corollary}[theorem]{Corollary}
\theoremstyle{definition}

\theoremstyle{remark}
\newtheorem{remark}[theorem]{Remark}
\newcommand{\Exp}{\mathbb E}
\newcommand{\Prob}{\mathbb P}
\newcommand{\ind}{\mathbf 1}
\newcommand{\Cov}{\text{\normalfont\scshape Cover-Time}}
\newcommand{\MaxCov}{\text{\normalfont\scshape Max-Cover-Time}}

\newcommand{\res}{R}
\title{How Hard Is It to Compute Cover Time?}
\author{Ryan Aghazadeh\\[0.3em]{\normalsize Cornell University}}
\date{}
\begin{document}
\vspace*{\stretch{1}}
\noindent\begin{minipage}{\textwidth}
\let\newpage\relax
\maketitle
\let\small\normalsize
\begin{abstract}
It is \#P-hard. Even for connected simple undirected unweighted graphs, deciding whether
an expected random-walk cover time is at most a given rational
threshold is \#P-hard under polynomial-time Turing reductions, both
from a specified starting vertex and when maximized over starting
vertices. Hardness holds even for graphs of diameter two with a universal vertex,
which we call the ``hub''.
Deleting the hub leaves a planar bipartite graph, and every non-hub vertex has degree at most five.
A shared construction encodes an Euler-tour count in rare covering walks.
For specified starts, a number of fractional bits linear in the queried
graph's vertex count suffices to recover the Euler-tour count. For the maximum-start problem,
a fixed cube attachment makes the maximizing starts known, and
one finite-difference calculation on expected cover times recovers the count.
For general graphs, both comparison problems can be solved in polynomial
time with access to an $\mathrm{NP}^{\mathrm{PP}}$ oracle.
\end{abstract}
\end{minipage}\par
\vspace*{\stretch{4}}
\clearpage

\section{Introduction}
Cover time measures how efficiently a random walk explores a graph.
For a finite connected graph, it is the first time the walk has visited every vertex. Exact expectations can be obtained by
expanding the state space to record the current vertex and the visited
set; this gives algorithms with exponential running time, including the approach of
Bui and Sohier~\cite{BuiSohier}. These expected cover times are rational, since they solve a linear system with rational coefficients~\cite{Sylvester}.

The complexity of computing expected cover time exactly has remained
unresolved. Georgakopoulos, Haslegrave, Sauerwald, and Sylvester explicitly posed this question in 2022
for the maximum expected cover time over starting vertices~\cite[p.~75]{GHSS}.
\textbf{We establish \#P-hardness} both for this maximum and for the expected
cover time from a specified starting vertex. Our results concern exact comparison with a
rational threshold, which may distinguish exponentially small additive
differences. Both quantities admit efficient approximation
algorithms~\cite{FeigeRabinovich,DingLeePeres}.

We define the decision problem $\Cov$ by
\begin{equation}\label{eq:language}
 \Cov=\{(G,s,r):C_G(s)\le r\},\qquad
 \text{where } C_G(s)=\Exp_s[\tau_G].
\end{equation}
Here $G$ is a connected simple undirected unweighted graph, $s$ is
its specified starting vertex, and $r$ is a binary-encoded rational
threshold. The walk chooses a neighbor uniformly. Its random cover time is
\[
 \tau_G=\min\{t\ge0:\{X_0,\ldots,X_t\}=V(G)\}.
\]
Graphs are given by neighbor lists, and rational thresholds by binary
integer pairs with positive denominator. The starting vertex is visited
at time zero, so a one-vertex graph has cover time zero.

We also define
\begin{equation}\label{eq:max-language}
 \MaxCov=\{(G,r):M(G)\le r\},\qquad
 \text{where } M(G)=\max_s C_G(s).
\end{equation}
The maximum-start language has the same graph and threshold encoding.

\begin{theorem}\label{thm:main}
$\Cov$ is \#P-hard under polynomial-time Turing reductions.

\smallskip
{\normalfont\leftskip=1.5em
\noindent The hardness holds even for graphs with a universal vertex $z$, called the hub,
and diameter two. Every non-hub vertex has degree at most five. Deleting
$z$ (and its incident edges) leaves a planar bipartite graph.\par}
\end{theorem}

\begin{theorem}\label{thm:max}
$\MaxCov$ is \#P-hard under polynomial-time Turing reductions.

\smallskip
{\normalfont\leftskip=1.5em
\noindent The hardness holds even for graphs with a universal vertex $z$, called the hub,
and diameter two. Every non-hub vertex has degree at most five. Deleting
$z$ (and its incident edges) leaves a planar bipartite graph.\par}
\end{theorem}

Counting Euler tours in general graphs is \#P-complete, as shown by
Brightwell and Winkler in 2005~\cite{BrightwellWinkler}.
Ge and \v Stefankovi\v c strengthened this result in 2012, proving
hardness even for planar 4-regular graphs~\cite{GeStefankovic}.
We will show that a polynomial-time algorithm with access to a $\Cov$
or $\MaxCov$ oracle can compute the exact number of Euler tours of a
connected planar 4-regular graph.

The specified-start reduction constructs one cover-time graph for each input graph.
The maximum-start reduction adds a cube and marker leaves, preserving
the same graph restrictions.

There is also a complementary upper bound for general graphs.
\begin{theorem}\label{thm:upper}
$\Cov$ and $\MaxCov$ are in
$\mathrm P^{\mathrm{NP}^{\mathrm{PP}}}$, and hence in the counting hierarchy.
\end{theorem}
Inclusion--exclusion expresses each expected cover time as an alternating
sum of expected hitting times, one for each nonempty vertex subset.
Each hitting time has a rational expectation computable in polynomial time.
Meiburg's comparison theorem~\cite{Meiburg} then gives the upper bound;
Appendix~\ref{sec:upper-bound} verifies its hypotheses. Here
$\mathrm P^{\mathrm{NP}^{\mathrm{PP}}}$ denotes deterministic polynomial
time with an $\mathrm{NP}^{\mathrm{PP}}$ oracle.

Our reductions also establish hardness of binary truncation.

\begin{corollary}[Hardness of binary truncation]\label{cor:prefix}
For $m=|V(G)|$, retaining the integer part and the first $2m$ bits after
the binary point of the expected cover time---that is, computing
\[
 \operatorname{Trunc}_{2m}(G,s)=\left\lfloor 2^{2m}C_G(s)\right\rfloor
\]
is \#P-hard under polynomial-time Turing reductions on the same
graph class of Theorem~\ref{thm:main}. On this class its output has only $O(m)$ bits.
More generally, the reduction works with rational approximations of
absolute error at most $2^{-2m}$.
\end{corollary}
Here the approximation oracle returns rationals of polynomial encoding
length. The accuracy requires only a linear number of fractional bits.

\paragraph{Proof structure.}
We reduce from counting Euler tours of a connected planar 4-regular
source graph $G$. Figure~\ref{fig:construction} shows the construction sequence.
Here $A$ is the anchored graph, $S$ is the subdivided graph, and $G'$
is the specified-start cover-time instance.
\begin{figure}[!t]
\centering
\begin{minipage}[t]{.47\linewidth}
\centering
\begin{tikzpicture}[x=2.25cm,y=1.8cm,
 edge/.style={line width=.7pt},
 vertex/.style={circle,fill=black,inner sep=1.7pt}]
\path[use as bounding box] (-1.35,-1.04) rectangle (1.35,1.22);
\draw[edge,line width=1.1pt] (-1,0)
 .. controls (-.55,1.15) and (.55,1.15) .. (1,0);
\draw[edge] (-1,0) .. controls (-.55,.4) and (.55,.4) .. (1,0);
\draw[edge] (-1,0) .. controls (-.55,-.4) and (.55,-.4) .. (1,0);
\draw[edge] (-1,0) .. controls (-.55,-1.15) and (.55,-1.15) .. (1,0);
\node[anchor=east,font=\small] at (-.2,1.14) {chosen edge $e^\ast$};
\draw[->,line width=.5pt] (-.16,1.1) -- (0,.88);
\node[vertex,label={left:$u$}] at (-1,0) {};
\node[vertex,label={right:$v$}] at (1,0) {};
\end{tikzpicture}
\par\smallskip
{\small (a) Source $G$: mark one of the four edges}
\end{minipage}
\hfill
\begin{minipage}[t]{.47\linewidth}
\centering
\begin{tikzpicture}[x=2.25cm,y=1.8cm,
 edge/.style={line width=.7pt},
 vertex/.style={circle,fill=black,inner sep=1.7pt}]
\path[use as bounding box] (-1.35,-1.04) rectangle (1.35,1.22);
\draw[edge,line width=1.1pt] (-1,0) .. controls (-.85,.42) and (-.6,.7) .. (-.36,.8);
\draw[edge,line width=1.1pt] (.36,.8) .. controls (.6,.7) and (.85,.42) .. (1,0);
\draw[edge] (-1,0) .. controls (-.55,.4) and (.55,.4) .. (1,0);
\draw[edge] (-1,0) .. controls (-.55,-.4) and (.55,-.4) .. (1,0);
\draw[edge] (-1,0) .. controls (-.55,-1.15) and (.55,-1.15) .. (1,0);
\node[vertex,label={left:$u$}] at (-1,0) {};
\node[vertex,label={right:$v$}] at (1,0) {};
\node[vertex,label={above left:$a$}] at (-.36,.8) {};
\node[vertex,label={above right:$b$}] at (.36,.8) {};
\end{tikzpicture}
\par\smallskip
{\small (b) Anchored $A$: replace $e^\ast$ by $au$ and $vb$}
\end{minipage}
\par\medskip
\begin{minipage}[t]{.49\linewidth}
\centering
\begin{tikzpicture}[x=2.25cm,y=1.8cm,
 vertex/.style={circle,fill=black,inner sep=1.5pt},
 internal/.style={circle,fill=black,inner sep=.85pt},
 checkpoint/.style={circle,draw=black,fill=white,inner sep=1.25pt,line width=.45pt},
 tube/.style={line width=.65pt}]
\path[use as bounding box] (-1.35,-1.35) rectangle (1.35,1.10);
\coordinate (u) at (-1,0);
\coordinate (v) at (1,0);
\coordinate (a) at (-.36,.8);
\coordinate (b) at (.36,.8);
\path (-1,0) .. controls (-.85,.42) and (-.6,.7) .. (-.36,.8)
 coordinate[pos=1/6] (left1) coordinate[pos=2/6] (left2) coordinate[pos=3/6] (left3) coordinate[pos=4/6] (left4) coordinate[pos=5/6] (left5);
\path (.36,.8) .. controls (.6,.7) and (.85,.42) .. (1,0)
 coordinate[pos=1/6] (right1) coordinate[pos=2/6] (right2) coordinate[pos=3/6] (right3) coordinate[pos=4/6] (right4) coordinate[pos=5/6] (right5);
\path (-1,0) .. controls (-.55,.4) and (.55,.4) .. (1,0)
 coordinate[pos=1/6] (upper1) coordinate[pos=2/6] (upper2) coordinate[pos=3/6] (upper3) coordinate[pos=4/6] (upper4) coordinate[pos=5/6] (upper5);
\path (-1,0) .. controls (-.55,-.4) and (.55,-.4) .. (1,0)
 coordinate[pos=1/6] (middle1) coordinate[pos=2/6] (middle2) coordinate[pos=3/6] (middle3) coordinate[pos=4/6] (middle4) coordinate[pos=5/6] (middle5);
\path (-1,0) .. controls (-.55,-1.15) and (.55,-1.15) .. (1,0)
 coordinate[pos=1/6] (lower1) coordinate[pos=2/6] (lower2) coordinate[pos=3/6] (lower3) coordinate[pos=4/6] (lower4) coordinate[pos=5/6] (lower5);
\draw[tube] (-1,0) .. controls (-.85,.42) and (-.6,.7) .. (-.36,.8);
\draw[tube] (.36,.8) .. controls (.6,.7) and (.85,.42) .. (1,0);
\draw[tube] (-1,0) .. controls (-.55,.4) and (.55,.4) .. (1,0);
\draw[tube] (-1,0) .. controls (-.55,-.4) and (.55,-.4) .. (1,0);
\draw[tube] (-1,0) .. controls (-.55,-1.15) and (.55,-1.15) .. (1,0);
\foreach \tube in {left,right,upper,middle,lower}{
 \foreach \i in {1,3,5}{\node[internal] at (\tube\i) {};}
 \foreach \i in {2,4}{\node[checkpoint] at (\tube\i) {};}
}
\node[vertex,label={left:$u$}] at (u) {};
\node[vertex,label={right:$v$}] at (v) {};
\node[vertex,label={above left:$a$}] at (a) {};
\node[vertex,label={above right:$b$}] at (b) {};
\draw[black!45,densely dotted,thin] (u) -- (-1.00000000,-0.40000000);
\draw[black!45,densely dotted,thin] (lower2) -- (-0.38148148,-1.01333333);
\draw[black!45,densely dotted,thin] (lower4) -- (0.38148148,-1.01333333);
\draw[black!45,densely dotted,thin] (v) -- (1.00000000,-0.40000000);
\draw[thin] (-1.00000000,-0.40000000) .. controls (-0.85000000,-0.70666667) and (-0.62777778,-0.91111111) .. (-0.38148148,-1.01333333)
 node[midway,sloped,below,font=\scriptsize] {$h$ edges};
\draw[thin] (-0.38148148,-1.01333333) .. controls (-0.13518519,-1.11555556) and (0.13518519,-1.11555556) .. (0.38148148,-1.01333333)
 node[midway,sloped,below,font=\scriptsize] {$h$ edges};
\draw[thin] (0.38148148,-1.01333333) .. controls (0.62777778,-0.91111111) and (0.85000000,-0.70666667) .. (1.00000000,-0.40000000)
 node[midway,sloped,below,font=\scriptsize] {$h$ edges};
\draw[thin] (-1.00000000,-0.43055556) -- (-1.00000000,-0.36944444);
\draw[thin] (-0.38148148,-1.04388889) -- (-0.38148148,-0.98277778);
\draw[thin] (0.38148148,-1.04388889) -- (0.38148148,-0.98277778);
\draw[thin] (1.00000000,-0.43055556) -- (1.00000000,-0.36944444);
\end{tikzpicture}
\par\smallskip
{\small (c) $S$: subdivide every edge of $A$}
\end{minipage}
\hfill
\begin{minipage}[t]{.47\linewidth}
\centering
\begin{tikzpicture}[x=1.9cm,y=1.05cm,
 vertex/.style={circle,fill=black,inner sep=1.5pt},
 internal/.style={circle,fill=black,inner sep=.85pt},
 checkpoint/.style={circle,draw=black,fill=white,inner sep=1.25pt,line width=.45pt},
 tube/.style={line width=.65pt}]
\path[use as bounding box] (-1.5,-1.65) rectangle (1.5,2.55);
\draw[rounded corners=3pt,black!60] (-1.35,-1.03) rectangle (1.35,1.18);
\node[anchor=north west] at (-1.32,1.17) {$S$};
\coordinate (hub) at (0,2);
\coordinate (u) at (-1,0);
\coordinate (v) at (1,0);
\coordinate (a) at (-.36,.8);
\coordinate (b) at (.36,.8);
\path (-1,0) .. controls (-.85,.42) and (-.6,.7) .. (-.36,.8)
 coordinate[pos=1/6] (left1) coordinate[pos=2/6] (left2) coordinate[pos=3/6] (left3) coordinate[pos=4/6] (left4) coordinate[pos=5/6] (left5);
\path (.36,.8) .. controls (.6,.7) and (.85,.42) .. (1,0)
 coordinate[pos=1/6] (right1) coordinate[pos=2/6] (right2) coordinate[pos=3/6] (right3) coordinate[pos=4/6] (right4) coordinate[pos=5/6] (right5);
\path (-1,0) .. controls (-.55,.4) and (.55,.4) .. (1,0)
 coordinate[pos=1/6] (upper1) coordinate[pos=2/6] (upper2) coordinate[pos=3/6] (upper3) coordinate[pos=4/6] (upper4) coordinate[pos=5/6] (upper5);
\path (-1,0) .. controls (-.55,-.4) and (.55,-.4) .. (1,0)
 coordinate[pos=1/6] (middle1) coordinate[pos=2/6] (middle2) coordinate[pos=3/6] (middle3) coordinate[pos=4/6] (middle4) coordinate[pos=5/6] (middle5);
\path (-1,0) .. controls (-.55,-1.15) and (.55,-1.15) .. (1,0)
 coordinate[pos=1/6] (lower1) coordinate[pos=2/6] (lower2) coordinate[pos=3/6] (lower3) coordinate[pos=4/6] (lower4) coordinate[pos=5/6] (lower5);
\foreach \point in {u,v,a,b}{\draw[black!25,line width=.35pt] (hub)--(\point);}
\foreach \tube in {left,right,upper,middle,lower}{
 \foreach \i in {1,...,5}{\draw[black!25,line width=.35pt] (hub)--(\tube\i);}
}
\draw[tube] (-1,0) .. controls (-.85,.42) and (-.6,.7) .. (-.36,.8);
\draw[tube] (.36,.8) .. controls (.6,.7) and (.85,.42) .. (1,0);
\draw[tube] (-1,0) .. controls (-.55,.4) and (.55,.4) .. (1,0);
\draw[tube] (-1,0) .. controls (-.55,-.4) and (.55,-.4) .. (1,0);
\draw[tube] (-1,0) .. controls (-.55,-1.15) and (.55,-1.15) .. (1,0);
\foreach \tube in {left,right,upper,middle,lower}{
 \foreach \i in {1,3,5}{\node[internal] at (\tube\i) {};}
 \foreach \i in {2,4}{\node[checkpoint] at (\tube\i) {};}
}
\node[vertex,label={left:$u$}] at (u) {};
\node[vertex,label={right:$v$}] at (v) {};
\node[vertex,label={above left:$a$}] at (a) {};
\node[vertex,label={above right:$b$}] at (b) {};
\node[vertex,label={above:$z$}] at (hub) {};
\end{tikzpicture}
\par\smallskip
{\small\makebox[\linewidth][r]{(d) $G'$: add the hub $z$ and join it to every vertex of $S$}}
\end{minipage}
\[
 G\xrightarrow{\substack{\text{cut the marked edge}\\\text{into leaf anchors}}}
 A\xrightarrow{\text{subdivide each edge}}
 S\xrightarrow{\text{add the hub }z}G'.
\]
\caption{The construction from $G$ to $G'$, illustrated with two vertices
joined by four distinct parallel edges. Cut the marked edge to form the
anchors of $A$, subdivide each edge to obtain $S$, then add the universal
hub $z$. The panels use illustrative block length $h=2$; in the reduction,
$h=64E$ and each tube has $3h=192E$ edges. Open circles mark internal
checkpoints; original vertices and anchors are also checkpoints. The
three blocks are marked on the bottom tube. Hub edges are light gray;
crossings of drawn edges are not vertices.}
\label{fig:construction}
\label{fig:tube-hub}
\end{figure}

\paragraph{Covering before the hub.}
Let's define the two types of walks we're interested in.

\begingroup
\leftskip=2em
\noindent\textbf{Type 1:} Starts at the hub, visits every vertex of $S$, and
ends before returning to the hub. The endpoint can be any time after
coverage---not necessarily the moment coverage occurs.

\noindent\textbf{Type 2:} Starts in $S$, visits every vertex of $S$, and ends
at its first visit to the hub.
\par
\endgroup

Reversing a walk sequence gives a bijection between these two types:
\[
 \underbrace{z,x_1,\ldots,x_s}_{\text{Type 1}}
 \quad\longleftrightarrow\quad
 \underbrace{x_s,\ldots,x_1,z}_{\text{Type 2}}.
\]
Counting Type 1 sequences along a walk measures the extra waiting time
between covering $S$ and returning to the hub, provided coverage occurs
first. Summing Type 2 probabilities measures the probability of covering
$S$ before reaching the hub. The path probabilities differ only by
endpoint-degree factors, relating this weighted probability to a difference
of expected cover times.

\paragraph{Recognizing Euler trails.}
Covering a tube does not require crossing it: a walk may visit its
vertices from both ends. Two internal checkpoints divide each tube into
three blocks; original vertices and anchors are also checkpoints. We
record successive visits to different checkpoints. Every tube contributes
two internal checkpoints that must be visited, and moving between tubes
requires an original vertex. The shortest recorded covering sequences
cannot afford any extra recorded visits: they necessarily follow Euler
trails between the two anchors of $A$. Every other Type 2 walk
requires an additional block crossing before coverage.

All Euler trails have the same departure counts at each type of
checkpoint, so local hitting probabilities give a common contribution
for one trail and its reverse. The same crossing probabilities appear
in the bounds for this contribution and for all the other walks together.
After division, the common crossing factors cancel. The probability of
an extra crossing decreases exponentially with block length, allowing
us to make the total error small enough for rounding.

\paragraph{Using maximum expected cover times.}
For the maximum-start problem, we attach a cube and marker leaves that
make the cube vertices the maximizing starts. Additional leaves reduce
the probability of choosing any particular neighbor at the hub, allowing
us to distinguish walks by their number of departures from it.
Truncating the sum over hub departures and clearing
denominators gives a polynomial whose leading coefficient is a known multiple of
the Euler-tour count, up to a small error.
\textbf{The key observation is that additional hub departures give lower-degree terms.}
The leading term therefore comes from walks with the fewest departures.
A finite difference extracts the coefficient, with
error small enough to recover the count by rounding.

\paragraph{Methodology and verification.}
Generative AI assisted the research and preparation of this paper, as
described in the AI methodology statement at the end.

The companion Lean development~\cite{CoverTimeLean} verifies both reductions end to end,
including their complexity bounds and the binary-truncation result.
We also independently verified and formalized Meiburg's comparison
theorem used for the upper bound. Euler-tour counting hardness and the
standard cover-time bound remain external assumptions. The companion
verification document records the precise statements, encodings, and
dependencies.

\section{Counting Euler tours and constructing the graph}\label{sec:construction}
Figure~\ref{fig:construction} illustrates the
construction: choose any edge of $G$, call it $e^*$, and cut it into two
leaf anchors; subdivide every edge, then add a universal hub.

Edges of $G$ have identities, and a self-loop has two distinct edge ends. We reserve $E(G)$ for the edge set and write $T(G)$ for the
number of Euler Tours. Tours are counted modulo cyclic rotation
and reversal, with self-loop incidence directions retained. Equivalently,
$T(G)$ counts the choices of pairings of the incident half-edges at
every vertex for which the edge links and pairing links form one global
circuit. There is no quotient by graph automorphisms.

\paragraph{Published hardness and counting conventions.}
Computing $T(G)$ on planar 4-regular sources is \#P-hard under
polynomial-time Turing reductions~\cite{BrightwellWinkler,GeStefankovic}.
The counting conventions agree as follows.

Brightwell and Winkler count the two directions of each Euler tour
separately, but do not distinguish starting points~\cite[Section 2]{BrightwellWinkler}.
Their count is therefore $2T(G)$ on the simple graphs used in their reduction.

Ge and \v Stefankovi\v c use the same counting convention, counting each
Euler tour once regardless of its starting point or direction~\cite[Sections 2--3]{GeStefankovic}.
Their bound $T(G)\le 3^{|V(G)|}$ follows because there are three ways to
pair the four edge ends at each vertex, and each Euler tour determines
one such pairing at every vertex.

Their Theorems 1 and 2 transfer the hardness to 4-regular and then
planar 4-regular instances. Our construction allows self-loops and parallel edges and applies to all
the graphs used in their reduction. We answer each Euler-tour counting query in their reduction by applying
our construction to the queried graph.
\paragraph{Cutting an edge to create anchors.}
Disconnected graphs have no Euler tours, and the one-vertex case can be counted directly.

We may therefore assume that $G$ is connected and has $n>1$ vertices.
Since $G$ is 4-regular, it has $2n$ edges. Choose any edge $e^\ast$
with endpoints $u,v$. Delete $e^\ast$, attach
a new leaf $a$ to $u$ and another new leaf $b$ to $v$, and call the
anchored graph $A$. If $e^\ast$ is a self-loop, then $u=v$ and the two distinct
leaves attach to the same vertex. Deleting an edge of a connected Eulerian graph
preserves connectivity: every edge lies on a cycle. The original
vertices of $A$ still have degree four, the leaves have degree one,
and
\begin{equation}\label{eq:E}
 E:=|E(A)|=2n+1.
\end{equation}
Cutting an Euler tour at $e^\ast$ and replacing that edge by the two leaf
edges gives a unique Euler \emph{trail} from $a$ to $b$. Conversely,
replacing the leaf edges by $e^\ast$ closes the trail into an Euler tour.
These operations are inverse, including when $e^\ast$ is a self-loop,
since its two edge ends are distinguished. Thus
\[
 \#\{\text{Euler trails in $A$ from $a$ to $b$}\}
 =T(G)
 =\#\{\text{Euler trails in $A$ from $b$ to $a$}\}.
\]

\paragraph{Subdividing the edges.}
We choose the block length large enough that the contribution from
non-Euler walks cannot affect rounding. Set
\begin{equation}\label{eq:length}
 h=64E.
\end{equation}
Replace each edge of $A$ by a path of $3h$ edges, called a
\textbf{\emph{tube}}, using distinct new internal vertices for different
edges. Call the resulting graph $S$.
A tube replacing a self-loop has both ends at the same vertex of $A$;
we distinguish its two ends when describing crossings and incursions.
A \textbf{\emph{crossing}} enters through one end and exits through the other;
an \textbf{\emph{incursion}} enters and exits through the same end.
\begingroup
\setlength{\abovedisplayskip}{6pt}
\setlength{\belowdisplayskip}{6pt}
\setlength{\abovedisplayshortskip}{6pt}
\setlength{\belowdisplayshortskip}{6pt}
\[
\text{Checkpoints}\quad
\left\{
\begin{aligned}
&\text{Vertices of the input graph and the two anchors},\\
&\text{Internal checkpoints: two per tube, at distances }h\text{ and }2h.
\end{aligned}
\right.
\]
\endgroup
These divide each tube into three blocks of $h$ edges. Designating
checkpoints does not change the graph or the random walk.

A self-loop becomes a cycle of length $3h$;
parallel edges become internally disjoint paths. Thus $S$ is simple,
even when $A$ has self-loops or parallel edges. The construction preserves planarity, and subdividing every edge into an
even-length path makes $S$ bipartite.
The subdivision adds $3h-1$ vertices per edge, so
$N=|V(S)|=n+2+E(3h-1)$. Since $n+2\le E$, we have
\begin{equation}\label{eq:N}
 E(3h-1)\le N\le3Eh=192E^2.
\end{equation}
\paragraph{Adding the hub.}
Add a vertex $z$ adjacent to every vertex of $S$, forming $G'$.
Every vertex is within two steps of every other, and the anchors $a,b$
are nonadjacent, so the diameter is exactly two. At every vertex of $S$,
the walk can move to $z$ on its next step. We will recover the Euler-tour
count from Type 2 walks.
\begin{center}
\begin{tabular}{lr}
\toprule
Vertex type & Degree in $G'$\\
\midrule
Original vertex of $G$ & $5$\\
Internal tube vertex & $3$\\
Leaf anchor $a$ or $b$ & $2$\\
Hub $z$ & $N=|V(S)|$\\
\bottomrule
\end{tabular}
\end{center}
All $N+1$ vertices are cover targets.

\section{A residual of expected full cover times}\label{sec:residual}
Reversing Type 1 walks into Type 2 walks relates expected cover times to
the probability of covering $S$ before reaching the hub. We first prove
the identity for an arbitrary vertex, since we will also use it to compare
starting vertices in the maximum-start reduction.

\begin{lemma}[Cover times and the last vertex discovered]\label{lem:last-vertex}
In a finite connected simple undirected graph with at least two vertices,
write $C_x$ for the expected cover time from $x$. For every vertex $v$,
\begingroup
\addtolength{\abovedisplayskip}{3.4pt}
\addtolength{\abovedisplayshortskip}{3.4pt}
\addtolength{\belowdisplayskip}{-8.7pt}
\addtolength{\belowdisplayshortskip}{-8.7pt}
\begin{equation}\label{eq:last-vertex}
 \deg(v)+\sum_{y\sim v}C_y-\deg(v)C_v
 =\sum_x\deg(x)\Prob_x(v\textup{ is the last vertex discovered}).
\end{equation}
\endgroup
\end{lemma}
\vspace*{-10.7pt}
\begin{proof}
Start the walk at $v$. Let $\tau$ be the first time it has visited every
other vertex, and let $H_v$ be its first return to $v$ after time zero.
Since $v$ is visited at time zero, the ordinary cover time is $\tau$.
If we ignore that initial visit, the walk must also return to $v$, so
coverage occurs at $\max(\tau,H_v)$. After the first step, the remaining
expected time is the cover time from a uniformly chosen neighbor. Hence
\begingroup
\addtolength{\abovedisplayskip}{-0.8pt}
\addtolength{\abovedisplayshortskip}{-0.8pt}
\addtolength{\belowdisplayskip}{-0.4pt}
\addtolength{\belowdisplayshortskip}{-0.4pt}
\[
 1+\frac1{\deg(v)}\sum_{y\sim v}C_y-C_v
 =\Exp_v[\max(H_v-\tau,0)].
\]
\endgroup
The right side is the expected waiting time from completing coverage until
the first return to $v$, with zero contribution if that return occurs before coverage.

For each time during this waiting period, reverse the walk sequence ending there.
A sequence $v=x_0,x_1,\ldots,x_s=x$ becomes a walk from $x$ whose
first visit to $v$ completes coverage. Conversely, every walk stopped
when it discovers $v$ last arises in this way. The probabilities satisfy
\[
 \Prob_v(X_0=x_0,\ldots,X_s=x_s)
 =\frac{\deg(x)}{\deg(v)}
   \Prob_x(X_0=x_s,\ldots,X_s=x_0),
\]
because the products of reciprocal degrees differ only at the two
endpoints. Summing these nonnegative terms and multiplying by $\deg(v)$
proves~\eqref{eq:last-vertex}. The same argument allows each sequence to carry a weight depending on
its final vertex; after reversal, that weight depends on its starting vertex.
\end{proof}

Apply Lemma~\ref{lem:last-vertex} to $G'$ at $v=z$, writing
$C_x=C_{G'}(x)$. Discovering $z$ last means covering $S$ before reaching it. The sequences counted before
reversal are Type 1 walks, and their reversals are Type 2 walks.
Since $\deg_{G'}(z)=N$, we obtain
\begin{equation}\label{eq:residual}
 \begingroup
 \setlength{\fboxsep}{6pt}
 \boxed{\begin{aligned}
 R&:=1+\frac1N\sum_{x\in S}C_x-C_z\\
  &\phantom{:}=\frac1N\sum_{x\in S}\deg_{G'}(x)
   \Prob_x(\text{cover }S\text{ before reaching }z).
 \end{aligned}}
 \endgroup
\end{equation}
Thus we can compute $R$ from expected cover times on $G'$. Each Type 2
walk starting at $x$ contributes its probability multiplied by
$\deg_{G'}(x)/N$; these weights need not sum to one.
All expectations here are finite because the graph is finite and connected.

\section{The walk observed at checkpoints}\label{sec:kernels}\label{sec:gap}
We now stop the walk on $G'$ when it first reaches $z$, also called
killing the walk at $z$.
Related reset and killed-walk methods appear in
Rivera, Sauerwald, and Sylvester~\cite{Rivera} and Si~\cite{Si}.

A walk can cover a tube from both ends without crossing it, or cross a
tube more than once. Recording visits to checkpoints distinguishes these
possibilities by the number of block crossings. The shortest routes covering
$S$ follow Euler trails and have exactly $3E$ block crossings; all other
covering routes have at least $3E+1$. Completing this additional crossing
before reaching the hub is exponentially unlikely in the block length.
We will use this to bound the error from non-Euler walks tightly enough
to recover the Euler-tour count by rounding.
\begin{figure}[H]
\centering
\begin{minipage}[t]{.47\linewidth}
\centering
\begin{tikzpicture}[x=.85cm,y=1cm,
 old/.style={circle,fill=black,inner sep=1.8pt},
 internal/.style={circle,fill=black,inner sep=1.4pt}]
\path[use as bounding box] (-.45,-.7) rectangle (6.45,1.5);
\draw[line width=.85pt] (0,0) -- (6,0);
\foreach \x in {1,3,5} \node[internal] at (\x,0) {};
\foreach \x in {2,4} \node[circle,draw,fill=white,inner sep=2.2pt] at (\x,0) {};
\node[old,label={below:$u$}] at (0,0) {};
\node[old,label={below:$v$}] at (6,0) {};
\draw[->,line width=.8pt] (.05,.5) -- (5.95,.5);
\node[font=\small,above] at (3,.6) {one complete crossing};
\end{tikzpicture}
\par\smallskip
{\small (a) Cross the tube}
\end{minipage}
\hfill
\begin{minipage}[t]{.47\linewidth}
\centering
\begin{tikzpicture}[x=.85cm,y=1cm,
 old/.style={circle,fill=black,inner sep=1.8pt},
 internal/.style={circle,fill=black,inner sep=1.4pt}]
\path[use as bounding box] (-.45,-.7) rectangle (6.45,1.5);
\draw[line width=.85pt] (0,0) -- (2,0);
\draw[dashed,black!45,line width=.85pt] (2,0) -- (3,0);
\draw[line width=.85pt] (3,0) -- (6,0);
\foreach \x in {1,3,5} \node[internal] at (\x,0) {};
\foreach \x in {2,4} \node[circle,draw,fill=white,inner sep=2.2pt] at (\x,0) {};
\node[old,label={below:$u$}] at (0,0) {};
\node[old,label={below:$v$}] at (6,0) {};
\draw[<->,line width=.8pt] (.05,.5) -- (1.95,.5);
\draw[<->,line width=.8pt] (3.05,.5) -- (5.95,.5);
\node[font=\small,above] at (1,.6) {incursion from $u$};
\node[font=\small,above] at (4.5,.6) {incursion from $v$};
\node[font=\small,below] at (2.5,-.2) {edge not traversed};
\end{tikzpicture}
\par\smallskip
{\small (b) Visit the tube from both ends}
\end{minipage}
\setlength{\belowcaptionskip}{-10.5pt}
\caption{Vertex coverage does not force a tube crossing. In panel (b),
an incursion from each end visits all five
internal vertices, although the dashed edge is never traversed.
Open circles mark the two internal checkpoints; the endpoints are also checkpoints. Both panels use illustrative
block length $h=2$. A complete crossing records three block crossings.
The two incursions shown on the right record four, even though neither
crosses the middle block.}
\label{fig:uncrossed-tube}
\end{figure}

\medskip
\noindent\begin{minipage}{\linewidth}
\begingroup
\leftskip=2em
\noindent\textbf{Route:} Record the first checkpoint reached, then each
subsequent visit to a different checkpoint, stopping when the walk reaches
the hub. The resulting sequence is the walk's route.

\noindent\hspace*{1em}\textbullet\enspace Notice that each step in the route corresponds to exactly one \textbf{block crossing}.
\par
\endgroup
\end{minipage}\par
\medskip

Returning to the current checkpoint does not add a block crossing; returning
to an earlier, different checkpoint does. For a finite initial walk
sequence, record only visits up to its endpoint. Route length counts
block crossings, not checkpoint visits.

\begin{lemma}[Shortest routes of covering walks]\label{lem:trace}
The route of any initial walk sequence covering $S$ before reaching the
hub has at least $3E$ block crossings. Equality holds exactly for routes
that follow an Euler trail between the anchors of $A$, crossing each
tube's three blocks in order. In that case, coverage first occurs on
arrival at the final anchor.
\end{lemma}
\begin{proof}
Group each tube's two internal checkpoints together, including its anchor when
it has one. There are $E$ groups containing $2E+2$ vertices altogether.
Every such route visits all of them. Moving from one group to another
requires passing through a vertex of the original graph $G$. Visiting
all $E$ groups therefore requires at least $E-1$ such intervening visits.
Thus the route has at least
\[
 \underbrace{2E+2}_{\text{internal checkpoint and anchor visits}}
 +\underbrace{E-1}_{\text{intervening original-vertex visits}}
 =3E+1
\]
checkpoint visits, hence at least $3E$ block crossings. We count
\emph{occurrences}, not distinct original vertices; the same original
vertex may be used repeatedly.

Equality leaves exactly $E-1$ original-vertex visits, all needed to move
between the $E$ groups. Revisiting a group would require another such
visit. Each group must therefore appear in one uninterrupted run, with
every internal checkpoint and anchor appearing exactly once.

Within each group, the checkpoints lie along one tube, so visiting each
exactly once forces the route to follow them in order. An anchor connects
to only one block, so visiting it in the middle of the route would require
visiting the checkpoint at the other end twice. The anchors must
therefore be the route's endpoints, and every tube is crossed exactly
once. This gives an Euler trail between the anchors of $A$. For a
self-loop, the order of its internal checkpoints distinguishes the two
directions.

Conversely, following an anchored Euler trail in this way gives a route
of $3E$ block crossings, visiting every vertex of $S$. The final anchor was not visited earlier, so coverage
first occurs on that final arrival.
\end{proof}

\paragraph{Crossing a block.}
After the first step into a block, let $p$ be the probability of reaching
the opposite endpoint before returning to the starting endpoint or
reaching the hub. All blocks have the same internal transition rule, so
$p$ is independent of the block and its orientation.
Taking $h-1$ consecutive forward steps succeeds with probability
$3^{-(h-1)}$. Conversely, success requires at least $h-1$ steps at internal
vertices, each avoiding the hub with probability $2/3$. Therefore
\begin{equation}\label{eq:block-probability}
 \boxed{3^{-(h-1)}\le p\le(2/3)^{h-1}.}
\end{equation}

At a checkpoint, the walk enters any given incident block with probability
at least $1/5$, so it crosses that block on its first attempt with probability
at least $p/5$.

For an upper bound, allow the walk to return to the checkpoint and try
again. Each attempt begins by choosing a block rather than the hub, with
probability at most $4/5$. Reaching the $j$th attempt requires making this
choice $j$ times, so its probability is at most $(4/5)^j$. The attempt
then succeeds with probability $p$. Summing over the attempts gives
\begin{equation}\label{eq:trace-step}
 \Prob(\text{next different checkpoint is reached before }z)
 \le p\sum_{j\ge1}(4/5)^j=4p.
\end{equation}
The bound allows failed returns through every incident block, including
ones used earlier in the route.

\section{Euler and non-Euler contributions}\label{sec:mainterm}
To identify the Euler contribution to $R$, let's define:

\begingroup
\leftskip=2em
\noindent\textbf{Euler walk:} A Type 2 walk (cover $S$ before first hitting the hub) whose first $3E$ block crossings
follow an Euler trail between the anchors of $A$. Only the route is
prescribed, the walk may backtrack within a block.

\noindent\textbf{\boldmath$F$:} The combined contribution to $R$ from Euler walks
following one fixed Euler trail between the anchors of $A$ and its reverse.
\par
\endgroup

By Lemma~\ref{lem:trace}, an Euler walk first covers $S$ upon reaching
the final anchor. We therefore need only calculate its contribution up
to that arrival, since later continuations have total probability one.

To obtain $F$, sum over starting vertices $x$, weighting the probability
of following either orientation by $\deg_{G'}(x)/N$, as in~\eqref{eq:residual}.
We will show that $F$ is independent of the chosen trail and can be
computed without counting Euler trails.

An Euler walk's route begins at an anchor, although the walk itself may
start inside the adjacent block. We compute the probability of:
\begin{list}{\arabic{enumi}.}{%
\usecounter{enumi}%
\setlength{\leftmargin}{2.5em}%
\setlength{\topsep}{0pt}%
\setlength{\partopsep}{0pt}%
\setlength{\itemsep}{0pt}%
\setlength{\parsep}{0pt}}
\item Reaching the initial anchor before any other checkpoint or the hub.
\item From each checkpoint, making the next block crossing along the given
route before reaching the hub.
\end{list}
Returns to the current checkpoint are allowed: the walk may enter an
incident block and come back without reaching another checkpoint, even if
that block was used earlier in the route. The probability of reaching the
next checkpoint depends only on the current checkpoint's degree in $G'$,
since all incident blocks have the same length and internal transition
probabilities.

\noindent\begin{minipage}{\linewidth}
Every anchored Euler route has the following departure counts:
\begin{center}
\begin{tabular}{lr}
\toprule
Checkpoint type & Total departures in the route\\
\midrule
Initial anchor & $1$\\
Internal checkpoint & $2E$\\
Original vertex of $G$ & $E-1$\\
\bottomrule
\end{tabular}
\end{center}
\end{minipage}

The last count is $2n=E-1$: the anchored Euler trail departs each
original vertex exactly twice, since it has degree four in $A$.

By the strong Markov property, multiply the weighted probability in part 1
by the transition probabilities in part 2, using the counts in the table.
The initial weights are $2/N$ at the anchor and $3/N$ at each internal
vertex of its adjacent block. The local probabilities and departure counts
are the same for every trail and its reverse, so they give the same product.
Doubling gives $F$.

Each calculation uses only one block or the blocks incident to one
checkpoint. Computing $F$ therefore requires neither tracking all visited
vertices nor counting Euler trails.

Each Euler walk follows one anchored Euler trail. Grouping these trails
with their reverses gives $T(G)$ pairs, each contributing $F$. Hence the
total Euler contribution to $R$ is $T(G)\cdot F$.

\paragraph{A lower bound on $F$.}
Retain only walks starting at the two anchors. Their combined weight is
$4/N$, and each of the $3E$ block crossings has probability at least $p/5$.
Hence
\begin{equation}\label{eq:mainterm}
 \boxed{F\ge\frac{4}{N}\left(\frac p5\right)^{3E}>0.}
\end{equation}
\subsection{Bounding the non-Euler contribution}\label{sec:mass}
By Lemma~\ref{lem:trace}, a non-Euler Type 2 walk makes at least $3E+1$
block crossings before covering $S$. We bound their total contribution by
including every walk that completes that many crossings before reaching
the hub, whether or not it covers $S$.

At each arrival at a checkpoint, the conditional probability of making
another block crossing before hitting the hub is at most $4p$ by
\eqref{eq:trace-step}. The strong Markov property therefore bounds the
probability of completing $q$ block crossings before the hub by $(4p)^q$;
the probability of reaching the first checkpoint is at most one.
The $N$ starting vertices each have weight at most $5/N$, so their weights
sum to at most five. It follows that
\begin{equation}\label{eq:source-remainder}
 0\le R-T(G)\cdot F\le5(4p)^{3E+1}.
\end{equation}
Combining~\eqref{eq:mainterm} and~\eqref{eq:source-remainder} bounds the
error after dividing by $F$:
\[
 \frac{R-T(G)\cdot F}{F}
 \le\frac{5(4p)^{3E+1}}{(4/N)(p/5)^{3E}}
 =5N20^{3E}p.
\]
The common factor $p^{3E}$ cancels, leaving one factor of $p$ for the
additional crossing. Its exponential decrease with the block length lets
us make the error small enough for rounding.

\begin{lemma}[Bounding the difference between $R/F$ and $T(G)$]\label{lem:witness}
For $h=64E$,
\begin{equation}\label{eq:rounding-gap}
 0\le\frac RF-T(G)<\frac1{32}.
\end{equation}
\end{lemma}
\begin{proof}
Since $h$ is even and $(2/3)^2<1/2$, the block bound gives
$p<2^{1-h/2}=2^{1-32E}$. Appendix~\ref{sec:witness-cover-proof}
combines this with $N\le192E^2$ to prove
\[
 5N20^{3E}p<\frac1{32}.
\]
\end{proof}

\section{Recovering the Euler-tour count}\label{sec:normalizer}
Recovering $T(G)$ requires dividing $R$ by $F$. We need to compute $F$
exactly and bound it below to control the resulting approximation error.

\begin{lemma}[Computing $F$ exactly and finding a lower bound]\label{lem:Fbits}
The coefficient $F$ is computable in polynomial bit time, has a
polynomial-bit rational representation, and satisfies
\begin{equation}\label{eq:F-dyadic-lower}
 4^N F>16.
\end{equation}
\end{lemma}
\begin{proof}
Use the two local calculations from Section~\ref{sec:mainterm}. Part 1
uses the block next to an anchor. For part 2, there is one calculation
for each checkpoint degree $2$, $3$, or $5$ in $G'$.

These hitting probabilities satisfy linear systems with $O(h)$
unknowns~\cite[Section 11.2]{GrinsteadSnell}. From every nonabsorbing state,
the next step reaches the hub with probability at least $1/5$, so
absorption occurs with probability one and the systems are nonsingular.
Exact rational elimination solves them in polynomial time~\cite{Bareiss}.
Combining the solutions with the weights and departure counts from
Section~\ref{sec:mainterm} gives $F$.

To bound the numerator and denominator of $F$, multiply each first-step
equation by the degree of its current vertex. Each row then has a constant
number of nonzero integer entries, all bounded in magnitude.
Hadamard's inequality bounds their determinants by $2^{O(h)}$;
Cramer's rule gives a common denominator of $O(h)$ bits for each system's
solutions.

Summing the initial values therefore retains $O(h+\log h)$
bits. There are only a fixed number of local systems and $3E$ transition
factors. Including the factor $1/N$ for the initial vertex, the resulting
product has $O(Eh+\log N)=O(N)$ bits.
This also proves polynomial bit complexity for forming the product.

Finally, combine the lower bound on $p$ in~\eqref{eq:block-probability} with the
lower bound on $F$ in~\eqref{eq:mainterm}:
\[
 \frac p5\ge\frac35\,3^{-h}>\frac12\,3^{-h},
 \qquad
 F>\frac{2^{1-3E-5Eh}}{N}.
\]
The bound on $F$ uses $3^3<2^5$. Appendix~\ref{sec:witness-cover-proof}
combines it with the bounds on $N$ to obtain \mbox{$4^NF>16$}.
\end{proof}

\subsection{Precision and oracle queries}\label{sec:precision}
To recover $T(G)$ by rounding, the combined error from non-Euler walks and
approximate cover times must be below $1/2$. We now choose a sufficient
accuracy and obtain it from threshold queries.

For $m=N+1$, the standard cover-time bound is
$2|E(G')|(m-1)$~\cite[Theorem 6.1]{AldousFill}.
Since $|E(G')|=3Eh+N\le3N$, we have
\begin{equation}\label{eq:cover-range}
 C_{G'}(x)\le6N^2<16m^2<2^m.
\end{equation}
The construction has $m\ge15$, so the last inequality holds. Thus
$[0,2^m]$ is a valid initial interval for bisection.

\begin{proposition}[Recovering the Euler-tour count from approximations]\label{prop:recovery}
Put $\epsilon=2^{-2m}$.
Suppose rational numbers
$\widehat C_x$, one for every vertex of $G'$, satisfy
\[
 |\widehat C_x-C_x|\le\epsilon.
\]
Define
\[
 \widehat{\res}=1+\frac1N\sum_{x\in S}\widehat C_x-\widehat C_z.
\]
Then
\begin{alignat}{2}
 \left|\frac{\widehat{\res}}F-T(G)\right|&<\frac18<\frac12,
 &\qquad T(G)&=\left\lfloor\frac{\widehat{\res}}F+\frac12\right\rfloor.
 \label{eq:recovery}
\end{alignat}
\end{proposition}
\begin{proof}
The average and the hub value each contribute at most $\epsilon$ to the
error, so $|\widehat{\res}-\res|\le2\epsilon$. Since
$\epsilon=4^{-(N+1)}$, Lemma~\ref{lem:Fbits} gives
$2\epsilon/F<1/32$. Together with~\eqref{eq:rounding-gap},
\[
 \left|\frac{\widehat{\res}}F-T(G)\right|
 \le\underbrace{\frac{|\widehat{\res}-\res|}{F}}_{\text{approximation error}}
 +\underbrace{\left|\frac\res F-T(G)\right|}_{\text{non-Euler contribution}}
 <\frac1{16}+\frac1{16}=\frac18.
\]
Nearest-integer rounding therefore recovers $T(G)$.
\end{proof}

\begin{proof}[\underline{Proof of Theorem~\ref{thm:main}}]
Given a connected planar 4-regular input graph $G$, construct $G'$ using
\eqref{eq:length} and compute $F$ by Lemma~\ref{lem:Fbits}.
Put $D=2^{2m}$.
For each starting vertex, begin with the bracket $[0,2^m]$ supplied by
Equation~\eqref{eq:cover-range}. Query $\Cov$ at its midpoint: a yes answer
replaces the upper endpoint, and a no answer replaces the lower endpoint.
The true expected cover time stays in the bracket, including at equality.
After $3m$ queries, the interval has width
\[
 \frac{2^m}{2^{3m}}=2^{-2m}=\frac1D.
\]
Return its upper endpoint as $\widehat C_x$, giving error at most $1/D$.
Proposition~\ref{prop:recovery} then recovers $T(G)$.

\textbf{Running time.} There are $m$ starting vertices and $3m$ queries per vertex, hence
exactly $3m\cdot m=3m^2$ comparison queries on this one graph.
All queried midpoints and returned endpoints are multiples of $1/D$.
Using denominator $D$, all endpoint and midpoint numerators are at most
$2^{3m}$, hence have at most $3m+1$ bits; $D$ has $2m+1$ bits.
The returned approximations share this denominator.
Computing $\widehat{\res}$, dividing by $F$, and rounding therefore take
polynomial time. By~\eqref{eq:N}, the graph has
\[
 \begin{aligned}
 |V(G')|&=m\le192E^2+1,\\
 |E(G')|&=3Eh+N\le384E^2.
 \end{aligned}
\]
Its labeled adjacency lists can be constructed in polynomial time.

\textbf{We have a polynomial-time Turing reduction from counting Euler tours
to $\Cov$ on the graph class of Theorem~\ref{thm:main}. Since Euler-tour
counting is \#P-hard, Theorem~\ref{thm:main} follows.}
\end{proof}

\begin{proof}[Proof of Corollary~\ref{cor:prefix}]
For each vertex $x$ of $G'$, obtain
$\pi_x=\lfloor 2^{2m}C_{G'}(x)\rfloor$ and set
$\widehat C_x=\pi_x/2^{2m}$. These approximations have error less than
$2^{-2m}$, so Proposition~\ref{prop:recovery} recovers $T(G)$.
One truncation query per vertex therefore suffices.
Equation~\eqref{eq:cover-range} gives $0\le\pi_x<2^{3m}$, so each
value has at most $3m$ bits.

The argument also applies to rational approximations with error at most
$2^{-2m}$ and polynomial encoding length. Even with different denominators,
computing $\widehat{\res}/F$ and rounding take polynomial time.
\end{proof}

\section{Attaching a cube to control the maximizing starts}\label{sec:helper}
For the maximum-start problem, we attach a cube and leaves to the hub
so that the cube vertices become the maximizing starts. The leaves make
a cube vertex unlikely to be discovered last. Lemma~\ref{lem:last-vertex}
converts this probability bound into a comparison with a hub start.
We bound starts in $S$ and at leaves separately.
\begin{figure}[H]
\centering
\begin{tikzpicture}[scale=.9,every node/.style={font=\small}]
\node[draw,ellipse,minimum width=2cm,minimum height=1.5cm] (Sregion) at (-3.5,0) {$S$};
\node[circle,fill,inner sep=2pt,label=below:$z$] (z) at (0,0) {};
\foreach \y in {-.35,.35}{\draw (z)--(-2.5,\y);}
\draw (z)--(Sregion.east);
\foreach \i/\x/\y in {1/1.7/-1.5,2/3.9/-1.5,3/3.9/.7,4/1.7/.7,5/2.5/-.7,6/4.7/-.7,7/4.7/1.5,8/2.5/1.5}{
 \coordinate (v\i) at ({\x+.425},\y);\draw[gray!55] (z)--(v\i);
}
\draw[thick] (v1)--(v2)--(v3)--(v4)--cycle;
\draw[thick] (v5)--(v6)--(v7)--(v8)--cycle;
\foreach \a/\b in {1/5,2/6,3/7,4/8}{\draw[thick] (v\a)--(v\b);}
\foreach \i in {1,...,8}{\fill (v\i) circle(2pt);}
\node[below] at (v2) {$c$};
\node[circle,fill,inner sep=2pt,label={[name=mfirst]above:$m_1$}] (m1) at (-.6,1.8) {};
\node[circle,fill,inner sep=2pt,label={[name=mlast]above:$m_N$}] (mN) at (.6,1.8) {};
\draw (z)--(m1);\draw (z)--(mN);\node at (0,1.8) {$\cdots$};
\draw[decorate,decoration={brace,amplitude=4pt}]
 ([yshift=2pt]mfirst.north west)--([yshift=2pt]mlast.north east)
 node[midway,above=6pt] {$N$ permanent markers};
\node[circle,fill,inner sep=2pt,label={[name=pfirst]below:$m_{N+1}$}] (p1) at (-.6,-1.8) {};
\node[circle,fill,inner sep=2pt,label={[name=plast]below:$m_j$}] (pj) at (.6,-1.8) {};
\draw (z)--(p1);\draw (z)--(pj);\node at (0,-1.8) {$\cdots$};
\draw[decorate,decoration={brace,mirror,amplitude=4pt}]
 ([yshift=-2pt]pfirst.south west)--([yshift=-2pt]plast.south east)
 node[midway,below=6pt] {$j-N$ probes};
\node at (-3.5,-1.2) {subdivided graph};
\end{tikzpicture}

\medskip
\begin{minipage}{\linewidth}
Let's define two types of \emph{leaves}:

\begingroup
\leftskip=2em
\noindent\textbf{Permanent markers:} The $N$ leaves attached to the hub in
 every query graph. Visiting them requires at least $N$ departures from the hub.

\noindent\textbf{Probes:} The additional leaves attached to the hub. Changing
 their number lets us separate walks by how many times they leave the hub.
\par
\endgroup
\end{minipage}
\caption{The cube helper in $Q_j$. The hub is joined to every vertex
of $S$, every cube vertex (gray edges), and every leaf $m_1,\ldots,m_j$.
The permanent markers are shown above the hub and the probes below;
when $j=N$, there are no probes.
For $j\ge N$, the maximizing starts are exactly the
eight cube vertices, with $c$ denoting an arbitrary cube vertex.
Edges within $S$ and most hub-to-$S$ edges are suppressed. Crossing lines
are not vertices.}
\label{fig:helper}
\end{figure}

\begin{samepage}
Add eight vertices indexed by three-bit strings, joining two when they
differ in one bit, and join all eight to $z$. Call this cube the helper.
For $j\ge N$, write $Q_j$ for the resulting graph with $j$ leaves at $z$: $N$ permanent
markers and $j-N$ probes. Thus $Q_N$ has no probes. The cube
has twelve edges and eight connections to the hub; each leaf adds one
vertex and one edge. Hence
\[
 \begin{aligned}
 |V(Q_j)|&=(N+1)+8+j=N+9+j,\\
 |E(Q_j)|&=|E(G')|+12+8+j=|E(G')|+20+j.
 \end{aligned}
\]
\end{samepage}
Fix any cube vertex, call it $c$. The hub remains universal. If it is removed, the remaining
graph consists of $S$, the planar bipartite cube, and isolated leaves.

\begin{lemma}[Known maximizing starts]\label{lem:helper}
For every $j\ge N$, the maximizing vertices of $Q_j$ are exactly its eight
cube vertices. Starting at any of them, the probability of covering the
cube before reaching the hub is a fixed computable rational
$\rho\ge2^{-14}$, independent of $j$.
\end{lemma}
\begin{proof}
\begin{samepage}
Throughout this proof write $C_x=C_{Q_j}(x)$.

\textbf{Starts outside the cube.} We first bound $C_x-C_z$ for $x\in S$.
The nonnegative right side of
\eqref{eq:last-vertex} gives the first-step inequality
\[
 C_x\le1+\frac1{\deg_{Q_j}(x)}\sum_{y\sim x}C_y.
\]
\end{samepage}
A step from $S$ stays in $S$ with probability at most $4/5$. Two
successive steps stay in $S$ with probability at most $8/15$: any
original vertex of $G$ has degree five and its neighbors other than
$z$ have degree three, while every other vertex of $S$ has degree at
most three. The two survival factors are therefore at most $4/5$ and
$2/3$, in one order or the other.

Put $\Delta=\max\{0,\max_{x\in S}(C_x-C_z)\}$. Subtract $C_z$ in
the first-step inequality and apply it once more at each neighbor in
$S$. A step to the hub has excess zero. The preceding one- and two-step
bounds give
\[
 \Delta\le1+\frac45+\frac8{15}\Delta,
 \qquad
 \Delta\le\frac{27}7<4-\frac18.
\]
If $\Delta=0$, the bound is immediate. At a leaf, the same first-step
inequality gives $C_x-C_z\le1$.

\textbf{Visiting a cube vertex.} Write $d=\deg_{Q_j}(z)=N+8+j$ and fix
a cube vertex $w$. An \textbf{excursion} starts at the hub and ends when
it next reaches the hub. We first show that an excursion visits $w$
with probability greater than $2/d$.

Take a walk from $w$ that first returns to $w$ before reaching $z$, and
redirect its final edge to $z$. This edge exists because the hub is
universal. The change preserves the walk's probability, since the
departing vertices are unchanged, and distinct walks give distinct
redirected walks. The direct step $w\to z$ is not among them, because
the graph has no self-loops. The two stopping events exhaust probability
one, so reaching $z$ before returning to $w$ has probability greater
than $1/2$.

Reversal multiplies these probabilities by $4/d$, as in
Lemma~\ref{lem:last-vertex}. Therefore
\[
 \begin{aligned}
 \Prob_z(\text{visit }w\text{ before returning to }z)
 &=\frac4d\,\Prob_w(\text{reach }z\text{ before returning to }w)\\
 &>\frac2d.
 \end{aligned}
\]

\begin{samepage}
\textbf{The last vertex discovered.} Suppose $r$ leaves and $w$ remain
unvisited when the walk is at the hub. An excursion visits a new leaf
with probability $r/d$ and visits $w$ with probability greater than
$2/d$. It cannot do both. Ignoring excursions that visit neither, the
next discovery among these vertices is a leaf with probability at most
\[
 \frac{r/d}{r/d+2/d}=\frac r{r+2}.
\]
\end{samepage}
After each new leaf, restart at the hub with one fewer unseen leaf. Thus
\[
 \Prob(\text{all $r$ unseen leaves precede }w)
 \le\prod_{i=1}^r\frac i{i+2}=\frac2{(r+1)(r+2)}.
\]

Before the first hub visit, the only leaf the walk can have visited is
its starting vertex. If $w$ was visited before the hub, it cannot be
last. Otherwise, at least $j-1$ leaves remain unseen. Applying the
preceding bound at the first hub visit gives, for every starting vertex $x$,
\[
 \Prob_x(w\text{ is the last vertex discovered})\le\frac2{j(j+1)}.
\]

\textbf{Comparing starting vertices.} Cube symmetry gives the same
expected cover time $C_c$ at all eight cube vertices. Since $c$ has three
cube neighbors and the hub, Lemma~\ref{lem:last-vertex} gives
\[
 4+C_z-C_c
 =\sum_x\deg_{Q_j}(x)
       \Prob_x(c\text{ is the last vertex discovered}).
\]
There are $d$ nonhub vertices, each of degree at most five, so the total
degree is at most $6d$. Since $j\ge N>288$ by~\eqref{eq:N}, we have $d=N+8+j\le3j$.
Consequently
\[
 0\le4+C_z-C_c
 \le\frac{12d}{j(j+1)}
 \le\frac{36}{j+1}<\frac18.
\]
Thus the excess over a hub start satisfies
\[
 C_x-C_z\quad
 \begin{cases}
 >4-1/8,&x\text{ is a helper vertex},\\[3pt]
 \le27/7<4-1/8,&\text{otherwise}.
 \end{cases}
\]
Precisely the cube vertices maximize, and $M(Q_j)=C_{Q_j}(c)$.

\textbf{Bounding and computing $\boldsymbol{\rho}$.} Before reaching the hub, a walk
started in the cube stays there, and its transition probabilities do
not depend on $j$. Following a fixed seven-step Hamiltonian path from
$c$ has probability $4^{-7}=2^{-14}$, so $\rho\ge2^{-14}$.

To compute $\rho$ exactly, record the current cube vertex and the set of
cube vertices visited, stopping at full cube coverage or the hub. This
gives a fixed finite system of rational first-step equations, which can
be solved exactly~\cite{Bareiss}.
\end{proof}

\section{Extracting the Euler count}\label{sec:helper-transform}
We now relate maximum expected cover times to the Type 2 walks used in
the specified-start reduction. The $N$ permanent markers require $N$
departures from the hub, and covering $S$ requires at least one more.
With exactly $N+1$ departures, $S$ must be covered in a single excursion.
Adding probes lets us isolate this contribution and recover the Euler-tour count.

Call $Q_N$ the base graph and put
\[
 D=\deg_{Q_N}(z)=2N+8.
\]
Let $X_s$ be the \textbf{base walk}, an ordinary random walk on $Q_N$
starting at $c$, let $\tau$ be its cover time, and let $r_s$ count the departures from the hub completed
by time $s$. The counter increases on a step out of the hub, and is
unchanged on a return to it. For every $s\ge\tau$, we have $r_s\ge N+1$.
Also $X_\tau\ne z$: the permanent markers require an earlier hub visit,
so the hub cannot be the last vertex discovered.

For $r\ge1$, once the $r$th departure is completed, $r_s=r$ until the
next departure. These positions include the nonhub part of one excursion
and its final hub visit. From any nonhub vertex, the next step reaches the
hub with probability at least $1/5$. Hence the expected number of nonhub positions in that
excursion is at most $\sum_{i\ge0}(4/5)^i=5$, followed by one hub position.
Restricting to positions after base coverage only decreases these counts:
\begin{equation}\label{eq:fixed-launch-mass}
 \begin{aligned}
 \Exp_c\!\left[\#\{s\ge\tau:r_s=r,\ X_s\ne z\}\right]&\le5,\\
 \Exp_c\!\left[\#\{s\ge\tau:r_s=r,\ X_s=z\}\right]&\le1.
 \end{aligned}
\end{equation}
These bounds justify the sums below and bound the terms we later discard.

We first combine the queried cover times so that contributions before
covering $Q_N$ cancel.

\begin{lemma}[A binomial sum of maximum expected cover times]\label{lem:helper-transform}
For $k\ge2$, put $x=D+k$. Then
\begin{equation}\label{eq:helper-transform}
 \begin{aligned}
 \operatorname{BinomialSum}(k)
 &:=
   \sum_{j=0}^k(-1)^{j+1}\binom kj M(Q_{N+j})\\[3pt]
 &\phantom{:}=\Exp_c\left[\sum_{s\ge\tau}
 \left\{\left(\frac D x\right)^{r_s}
 -\frac{2k}{x-1}\cdot\ind_{\{X_s=z\}}
       \left(\frac D{x-1}\right)^{r_s}\right\}\right].
 \end{aligned}
\end{equation}
\end{lemma}
\begin{proof}
Run one walk from $c$ on the graph with all $k$ probes. For each subset
$I$ of those probes, erase every two-step loop through a probe outside
$I$: delete the probe position and its following hub-return position,
retaining the preceding hub position. Call the resulting sequence the
projected walk for $I$. Its departures from the hub are uniform among
the $D+|I|$ retained neighbors, so it is an ordinary walk on the graph
retaining exactly those probes. Its expected cover time is $M(Q_{N+|I|})$, by
Lemma~\ref{lem:helper} and symmetry among the probes.

Each projected cover time counts retained positions strictly before
that projected walk has visited every required vertex. Sum the projected
cover times with sign $(-1)^{|I|+1}$. There are $\binom{k}{j}$ subsets of
size $j$, so the expectation is the binomial sum in~\eqref{eq:helper-transform}.
Every projected cover time is
bounded by the full walk's cover time, so this finite signed sum is
integrable and we may cancel its contributions position by position.

\textbf{Cancellation.} Before the base graph is covered, every projection is still uncovered.
A base position is retained for every $I$. A probe position and its
following hub return are retained exactly when $I$ contains that probe.
Both signed sums are zero, since $k\ge2$.

\begin{samepage}
After base coverage, let $U$ be the set of probes already visited,
including the current arrival. A retained position contributes to the
projected cover time exactly when some probe in $I$ remains unvisited,
that is, $I\not\subseteq U$. Alternating over $I$ leaves
only the following contributions:
\begin{center}
\begin{tabular}{p{.48\linewidth}p{.42\linewidth}}
\toprule
Position in the full walk & Contribution after cancellation\\
\midrule
A nonhub base vertex, or a hub arrival from a base vertex
 & $+1$ if no probe has yet been visited; otherwise zero\\[4pt]
A probe, or its following hub return
 & $-1$ if this is the only probe visited so far; otherwise zero\\
\bottomrule
\end{tabular}
\end{center}
\end{samepage}
For the first row, subtract the signed sum over $I\subseteq U$ from
the sum over all $I$. It is nonzero only when $U$ is empty, giving $+1$.
For the second row, require $I$ to contain the current probe. The sum
over such subsets of $U$ is nonzero only when $U$ consists of that
probe alone, giving $-1$ after subtraction.

\textbf{Evaluating the two terms.} Erase all probe loops to obtain the base walk $X_s$. For a base position
with $r_s$ departures, conditional on the base walk, the probability
that all those departures occur without a probe visit is $(D/x)^{r_s}$. This gives the first term
in~\eqref{eq:helper-transform}, including the hub arrivals from base vertices.

For the negative contribution, fix the sole probe that may have been
visited and avoid the other $k-1$ probes. After erasing loops through
the allowed probe, each departure chooses among the $D$ base neighbors
and the $k-1$ forbidden probes. Reaching a given base position without
visiting a forbidden probe therefore contributes $(D/(x-1))^{r_s}$.

At a hub position, the expected number of loops through the allowed
probe before any other departure is $\sum_{i\ge1}x^{-i}=1/(x-1)$.
Each such loop contributes two positions with value $-1$. Summing over
the $k$ possible probes gives the factor $-2k/(x-1)$, which appears only
at hub positions. The allowed probe may have appeared before
base coverage; its earlier loops are erased as well, and each counted
loop ensures that it has now been visited.

For each fixed $r$, the expected number of all base positions is at
most six and of hub positions at most one, by~\eqref{eq:fixed-launch-mass}.
Both ratios $D/x$ and $D/(x-1)$ are less than one because $k\ge2$.
The resulting geometric bounds prove absolute integrability of the two
sums and justify taking their expectations separately.
\end{proof}

\paragraph{Isolating the leading coefficient.}
We truncate the sum in~\eqref{eq:helper-transform} and clear its denominators
so that positions with exactly $N+1$ departures determine the leading
coefficient. Set
\[
 K=1024D,\qquad t=2K-N.
\]
For the moment retain only the positions with $r_s\le K$ in
\eqref{eq:helper-transform}. Multiplying by $x^K(x-1)^{K+1}$ removes both
denominators. A position with $r_s=r$ then contributes
\begin{equation}\label{eq:cleared-position}
 D^r\left[x^{K-r}(x-1)^{K+1}
 -2(x-D)x^K(x-1)^{K-r}\cdot\ind_{\{X_s=z\}}\right].
\end{equation}
Its degree is at most $2K+1-r$, and its coefficient at that degree is
\[
 \begin{cases}
 D^r,&X_s\ne z,\\
 -D^r,&X_s=z.
 \end{cases}
\]
Since $r_s\ge N+1$ after coverage, the degree is at most $t$.
Each additional departure lowers the degree by one, so only positions
with exactly $N+1$ departures can contribute to the coefficient of $x^t$.
The finite range of departure counts and
\eqref{eq:fixed-launch-mass} justify summing these coefficients and
taking expectations.

Apart from the common factor $D^r$, subtracting the final hub position
leaves the number of steps between nonhub positions after coverage.
In particular, a marker visit and its following hub return cancel.

\textbf{The contribution with $\boldsymbol{N+1}$ departures.}
Exactly $N+1$ departures allow one visit to each permanent marker and
one excursion into $S$. No departure remains for another visit to the
cube, so it must already have been covered before the first hub visit,
an event of probability $\rho$. If the excursion
into $S$ is not the last of these excursions, base coverage ends at a
marker and the contribution is zero. Otherwise the $N!$ marker orders
contribute $N!/D^N$, and the final departure enters $S$ with probability
$N/D$. Conditional on entering $S$, its first vertex is uniform on $S$,
and the walk until its next hub hit is exactly the killed walk used
earlier. Thus the combined factor is
\[
 \rho\cdot\frac{N!}{D^N}\cdot\frac ND\cdot D^{N+1}=\rho N N!.
\]
The remaining factor is the expected number of steps within $S$ after
covering $S$ and before reaching the hub, with a uniform initial vertex.

To count only steps that stay in $S$, weight each position after coverage
by $1-1/\deg_{G'}(X_s)$. Applying the weighted reversal identity from
Lemma~\ref{lem:last-vertex} moves this weight to the starting vertex $x$
and multiplies it by $\deg_{G'}(x)$, giving $\deg_{G'}(x)-1$.
Thus the degree-$t$ coefficient is
\begin{equation}\label{eq:max-source-probability}
 \rho N N!\cdot\frac1N\sum_{x\in S}
 \bigl(\deg_{G'}(x)-1\bigr)
 \Prob_x(\text{cover }S\text{ before reaching }z).
\end{equation}
This sums over the same Type 2 walks as in~\eqref{eq:residual}, with initial
weight $(\deg_{G'}(x)-1)/N$ instead of $\deg_{G'}(x)/N$.

\paragraph{The exact coefficient $\boldsymbol{Z}$.}
Let $Z$ be the contribution to~\eqref{eq:max-source-probability} from Euler
walks following one fixed anchored trail and its reverse. Compute it as
for $F$, changing the initial weights to $1/N$ at the anchor and $2/N$
inside its adjacent block, then multiplying by $\rho N N!$. The local
probabilities and the factor of two for the orientations are unchanged.

The new initial weights are at least half the old ones and never larger.
Euler walks therefore contribute $T(G)\cdot Z$ to~\eqref{eq:max-source-probability},
while the remaining contribution is nonnegative and at most
$\rho N N!\,(R-T(G)\cdot F)$. The bounds on $F$ and $R-T(G)\cdot F$
therefore give
\begin{equation}\label{eq:max-normalizer}
 \begin{aligned}
 Z&\ge\frac{\rho N N!}{2}F>2^{-D}N!,\\
 0&\le\frac{\text{degree-$t$ coefficient}}{Z}-T(G)<\frac1{16}.
 \end{aligned}
\end{equation}
For the first line, use $4^NF>16$, $\rho\ge2^{-14}$, and $D=2N+8$:
the resulting lower bound is $(N/8)2^{-D}N!$, which exceeds $2^{-D}N!$.
For the second, after dividing by $Z$, the non-Euler contribution is
at most $2(R-T(G)\cdot F)/F<1/16$.

The local systems used for $F$ also compute $Z$ in polynomial time.
Its rational encoding uses $O(N\log N)$ bits, including the factorial.

\subsection{Finite differences and rounding}\label{sec:max-composition}
We extract the coefficient of $x^t$ using a $t$th finite difference.
This cancels all lower-degree terms and multiplies that coefficient by
$t!$, so dividing by $Zt!$ gives $T(G)$ up to the error
in~\eqref{eq:max-normalizer}. Set
\[
 x_a=2K+a,\qquad k_a=x_a-D\qquad(0\le a\le t).
\]
Approximate each required maximum expected cover time to error at most
$2^{-24K}$, and form the corresponding binomial sums
$\widehat{\operatorname{BinomialSum}}(k_a)$. Define
\begin{equation}\label{eq:max-decoder}
 \widehat T=
 \frac1{Zt!}\sum_{a=0}^t(-1)^{t-a}\binom ta
 x_a^K(x_a-1)^{K+1}\,
 \widehat{\operatorname{BinomialSum}}(k_a).
\end{equation}
The full binomial sums still include terms with more than $K$ departures
from the hub, and the cover-time values are approximate.
Appendix~\ref{sec:helper-precision} bounds both resulting errors.
Together with the non-Euler bound in~\eqref{eq:max-normalizer}, this gives
$|\widehat T-T(G)|<3/32$.

\begin{proof}[\underline{Proof of Theorem~\ref{thm:max}}]
Construct $G'$, attach the cube and permanent markers, and compute $Z$
by the local calculation above. Lemma~\ref{lem:helper} identifies a
maximizing start in every queried graph. Use the $\MaxCov$ oracle to
approximate the required values $M(Q_j)$ to error at most $2^{-24K}$ by
bisection, then compute~\eqref{eq:max-decoder}. Then
\[
 \left|\widehat T-T(G)\right|<\frac3{32}<\frac12,
 \qquad T(G)=\left\lfloor\widehat T+\frac12\right\rfloor.
\]
Nearest-integer recovery only requires error strictly below $1/2$.
Only $O(N)$ distinct cover-time values are needed across all the binomial
sums. Appendix~\ref{sec:helper-precision} shows that each requires $O(N)$
threshold queries, for $O(N^2)$ queries in total, and that the arithmetic
takes polynomial time.

Every query graph is $Q_j$ with $j=O(N)$, so it has $O(N)=O(E^2)$
vertices and edges. Deleting $z$ leaves $S$, a disjoint cube, and isolated
leaves, hence a planar bipartite graph. Vertices of $S$ retain their
degrees from $G'$; cube vertices have degree four and leaves degree one,
so every vertex other than $z$ has degree at most five. The universal
hub and the nonadjacent anchors give diameter exactly two.

\textbf{We have a polynomial-time Turing reduction from counting Euler tours
to \textnormal{\scshape Max-Cover-Time} on the graph class of Theorem~\ref{thm:max}. Since Euler-tour
counting is \#P-hard, Theorem~\ref{thm:max} follows.}
\end{proof}

\section{Scope and further questions}\label{sec:scope}
Theorems~\ref{thm:main}--\ref{thm:upper} give the oracle bounds
\[
 \begin{array}{c@{\;}c@{\;}c@{\;}c@{\;}l}
 \mathrm P^{\mathrm{PP}}&\subseteq&\mathrm P^{\Cov}
                    &\subseteq&\mathrm P^{\mathrm{NP}^{\mathrm{PP}}},\\[3pt]
 \mathrm P^{\mathrm{PP}}&\subseteq&\mathrm P^{\MaxCov}
                    &\subseteq&\mathrm P^{\mathrm{NP}^{\mathrm{PP}}}.
 \end{array}
\]
The left inclusions follow because either cover-time oracle can recover
\#P counts and therefore answer PP queries. The right inclusions follow
from Theorem~\ref{thm:upper}. The gap between the bounds remains open.
Our hardness results use Turing reductions and do not establish hardness
under many-one reductions.

On connected unweighted graphs, Monte Carlo simulation approximates a
specified-start expected cover time to inverse-polynomial additive error
in randomized polynomial time~\cite{FeigeRabinovich}.
The polynomial upper bound on expected cover time lets us obtain this
accuracy by requesting sufficiently small relative
error~\cite[Theorem 6.1 and Section 6.8.3]{AldousFill}.
Corollary~\ref{cor:prefix} establishes hardness at exponentially finer
additive accuracy.

The hub has unbounded degree. Although $G'-z$ is both planar and bipartite,
$G'$ has triangles, and the theorem does not impose planarity on $G'$.
Hardness with every degree bounded and hardness on trees remain natural
questions.

\paragraph{Related work.}
Ball, Dunham and Hirschowitz give exact formulas for cover-time generating
functions and moments~\cite{Ball}. Sericola develops exact methods for
cover-time distributions and moments~\cite{Sericola}.

Rivera, Sauerwald, and Sylvester add a universal vertex with weighted edges
so that visits to it reset the walk to the original stationary
distribution~\cite[Definition 3.7]{Rivera}. Our construction uses unweighted
edges, and the analysis stops the walk at its first hub visit.
Si uses killed Green matrices and inclusion--exclusion to compute the
exact change in expected cover time when an edge is added, for a specified
starting vertex~\cite{Si}.

\appendix
\section{Error and precision bounds for the specified-start reduction}\label{sec:witness-cover-proof}
We complete the numerical bounds in Lemmas~\ref{lem:witness} and~\ref{lem:Fbits}.
Both use $h=64E$ and
\[
 E(3h-1)\le N\le3Eh=192E^2.
\]
For $E\ge1$, $E^2\le2^{2E-2}$: it holds at one, and
$(E+1)^2\le4E^2$ preserves it. Consequently
\begin{equation}\label{eq:source-size-certificate}
 N\le192E^2\le48\cdot2^{2E}<2^{2E+6}.
\end{equation}

\paragraph{Non-Euler contribution.}
Section~\ref{sec:mass} gives $(R-T(G)\cdot F)/F\le5N20^{3E}p$.
Using~\eqref{eq:source-size-certificate}, $5<2^3$, $20<2^5$, and
$p<2^{1-32E}$, we obtain
\[
 \frac{R-T(G)\cdot F}{F}
 \le5N\,20^{3E}p
 <2^3\cdot2^{2E+6}\cdot2^{15E}\cdot2^{1-32E}
 =2^{10-15E}<\frac1{32},
\]
where $E\ge5$ under the source convention of Section~\ref{sec:construction}.

\paragraph{Precision.}
The estimate for $F$ in the proof of Lemma~\ref{lem:Fbits} and
$N\ge E(3h-1)$ give
\[
 4^NF>\frac{2^{1+Eh-5E}}N
       >2^{Eh-7E-5}>16.
\]
The middle inequality reuses~\eqref{eq:source-size-certificate}, and the
last follows from $h=64E$ and $E\ge5$.

\section{Error and precision bounds for the maximum-start reduction}\label{sec:helper-precision}
\begin{samepage}
We bound the error in~\eqref{eq:max-decoder} and show that evaluating it
requires polynomial-time arithmetic and $O(N^2)$ threshold queries. Recall
\[
 \begin{gathered}
 D=2N+8,\qquad K=1024D,\qquad t=2K-N,\\
 x_a=2K+a,\qquad k_a=x_a-D\qquad(0\le a\le t).
 \end{gathered}
\]
All chosen values satisfy $2D+2\le x_a<4K$.
\end{samepage}

\paragraph{Retained terms.}
For the retained positions $r_s\le K$, the degree calculation in
Section~\ref{sec:helper-transform} shows that the $t$th finite difference
extracts $t!$ times the degree-$t$ coefficient. This operation passes
through the sum and expectation: there are finitely many retained
departure counts and evaluation nodes, and~\eqref{eq:fixed-launch-mass}
bounds the expected number of positions for each count by six. At each
node, each position's contribution has a fixed bound. After division by
$Zt!$, \eqref{eq:max-normalizer} therefore bounds the error from the
retained terms by $1/16$.

\paragraph{Discarded terms.}
At any selected value $x$, the first term in~\eqref{eq:helper-transform}
with fixed departure count $r$ has expected absolute contribution at
most $6(D/(x-1))^r$. The second has expected absolute contribution at
most $2(D/(x-1))^r$: there is at most one hub position and
$2(x-D)/(x-1)\le2$. After clearing denominators, the absolute contribution
from all omitted counts is therefore at most
\begin{align*}
 &8x^K(x-1)^{K+1}
       \sum_{r>K}\left(\frac D{x-1}\right)^r\\
 &\qquad=\frac{8D^{K+1}x^K}{1-D/(x-1)}
       \le16D^{K+1}x^K.
\end{align*}
The last step uses $D/(x-1)\le1/2$.

\paragraph{Approximation error.}
Each queried maximum expected cover time is approximated with error at
most $2^{-24K}$, so each binomial sum has error at most $2^{k-24K}$.
After clearing denominators, this error is at most
\[
 D^{K+1}x^K
 \left(\frac{x-1}{D}\right)^{K+1}2^{k-24K}
 <D^{K+1}x^K\,2^{12-8K}
 \le D^{K+1}x^K.
\]
Here $x-1<4K=2^{12}D$ and $k=x-D<4K$. The discarded terms and
approximation therefore contribute a total error at node $x_a$ of less than
\begin{equation}\label{eq:maximum-pointwise-error}
 16D^{K+1}x_a^K+D^{K+1}x_a^K<64D^{K+1}x_a^K.
\end{equation}
This absolute-error bound remains valid when an approximation is reused
in several binomial sums.

\paragraph{Combined error.}
Apply the finite-difference formula in~\eqref{eq:max-decoder}, including
division by $Zt!$, to the errors bounded in~\eqref{eq:maximum-pointwise-error}.
The binomial coefficients have absolute sum $2^t$, and $x_a<4K$.
The combined error after division by $Zt!$ is therefore at most
\[
 \frac{64\cdot2^tD^{K+1}(4K)^K}{Zt!}.
\]
To bound it, use $N+t=2K$ and compare $K^{2K}$ with the product of the
$2K$ integers $K,K+1,\ldots,3K-1$:
\[
 \frac{K^{2K}}{N!t!}
 \le\binom{2K}{N}\binom{3K-1}{2K}
 \le2^{5K-1}.
\]
Combining this with $Z>2^{-D}N!$ and $t<2K$ gives
\[
 \frac{64\cdot2^tD^{K+1}(4K)^K}{Zt!}
 <2^{D+5}D\left(\frac{512D}{K}\right)^K
 =2^{D+5}D\,2^{-K}
 \le2^{3D-K}<\frac1{64}.
\]
We used $D\le2^D$, $D\ge5$, and $K=1024D$. Adding the non-Euler
contribution below $1/16$ yields
\[
 |\widehat T-T(G)|<\frac1{16}+\frac1{64}<\frac3{32},
\]
which proves the asserted rounding margin.

\paragraph{Arithmetic and oracle queries.}
Write $Z=u/v$ in lowest terms. Expanding the two finite sums in
\eqref{eq:max-decoder} gives rational coefficients with common denominator
$u\,t!$. Their numerators are sums of products of $v$, binomial
coefficients with $O(N)$ upper indices, and integer powers with $O(N)$
exponents and $O(N)$ bases. Each product has $O(N\log N)$ bits, and
summing polynomially many terms adds only $O(\log N)$ bits. All
coefficients are therefore constructible in polynomial bit time and
have $O(N\log N)$-bit encodings.

Only the $O(N)$ graphs $Q_{N+j}$, $0\le j\le k_t<4K$, are needed;
overlapping binomial sums reuse the same values. Each graph has $O(N)$
vertices and edges, so the standard cover bound~\cite{AldousFill} gives
an initial interval $[0,2^b]$ with $b=O(\log N)$.
The bisection procedure of Section~\ref{sec:precision} reaches width
$2^{-24K}$ after $b+24K=O(N)$ queries. Thus $O(N^2)$ threshold queries
suffice. Returning the upper endpoints gives the required accuracy,
including at equality, with common denominator $2^{24K}$. This adds
only $O(N)$ bits to the coefficient arithmetic, so evaluating
\eqref{eq:max-decoder} and rounding take polynomial time.
\begin{center}
\begin{tabular}{ll}
\toprule
Resource & Sufficient bound in terms of $N=|V(S)|$\\
\midrule
Vertices and edges per maximum-start query & $O(N)=O(E^2)$\\
Accuracy bits per queried expected cover time & $O(N)$\\
Number of queried maximum expected cover times & $O(N)$\\
Threshold comparisons & $O(N^2)$\\
Bits needed to represent each exact coefficient & $O(N\log N)$\\
\bottomrule
\end{tabular}
\end{center}

\section{A counting-hierarchy upper bound}\label{sec:upper-bound}
We write expected cover time as an alternating sum of expected hitting
times. Each term can be computed in polynomial time, and Meiburg's
theorem lets us compare the sum with the threshold.

\begingroup
\setlength{\abovedisplayskip}{5pt plus 1pt minus 1pt}
\setlength{\belowdisplayskip}{5pt plus 1pt minus 1pt}
\setlength{\abovedisplayshortskip}{0pt plus 1pt}
\setlength{\belowdisplayshortskip}{5pt plus 1pt minus 1pt}
\makeatletter
\renewcommand{\paragraph}{\@startsection{paragraph}{4}{\z@}%
  {5pt plus 1pt minus 1pt}{-1em}{\normalfont\normalsize\bfseries}}
\renewcommand{\section}{\@startsection{section}{1}{\z@}%
  {7pt plus 1pt minus 1pt}{6pt plus 1pt}{\normalfont\Large\bfseries}}
\makeatother
\begin{proof}[\underline{Proof of Theorem~\ref{thm:upper}}]
\begin{samepage}
\leavevmode
\paragraph{Expressing cover times as a sum.}
For nonempty $W\subseteq V(G)\setminus\{s\}$, let $H_W$ be its
first hitting time. Inclusion--exclusion of the unvisited-vertex events,
followed by the tail-sum formula, gives
\begin{equation}\label{eq:hit-set-sum}
 C_G(s)=\sum_{\varnothing\ne W\subseteq V(G)\setminus\{s\}}
       (-1)^{|W|+1}\Exp_s[H_W].
\end{equation}
The finite subset sum and finite expectations justify the interchange.
When $G$ has one vertex, this is an empty sum and $C_G(s)=0$.
\end{samepage}

For each nonempty $W$, make its vertices absorbing. Connectivity gives
a unique solution to the first-step equations~\cite[Section 11.2,
Theorems 11.4--11.5]{GrinsteadSnell}
\[
 t(x)=1+\frac1{\deg_G(x)}\sum_{y\sim x}t(y)\quad(x\notin W),
 \qquad t(x)=0\quad(x\in W).
\]
Here $t(s)=\Exp_s[H_W]$. The system has at most $n=|V(G)|$ unknowns
and polynomial-bit rational entries. Exact rational elimination computes
its solution in polynomial bit time, with polynomial-bit numerators and
denominators~\cite{Bareiss}. Thus the $(n-1)$-bit indicator of $W$
determines a polynomial-time computable signed rational summand
in~\eqref{eq:hit-set-sum}.

\paragraph{Comparing with the threshold.}
Meiburg's theorem applies to sums of polynomial-time computable rational
weights indexed by binary strings of polynomial length. Each numerator and
positive denominator must have polynomial bit length; the denominator
of the entire sum need not. It places comparison of such a sum with a
rational threshold in $\mathrm P^{\mathrm{NP}^{\mathrm{PP}}}$,
including equality~\cite[Appendix K, Theorem 12]{Meiburg}.

The summands in~\eqref{eq:hit-set-sum} satisfy these hypotheses.
The theorem is stated
for comparison $\ge$; negating the weights and the threshold tests
$-C_G(s)\ge-r$, which is exactly $C_G(s)\le r$.
For $\MaxCov$, perform this comparison for every starting vertex and
accept exactly when all answers are yes. Polynomially many such
comparisons remain in the same deterministic oracle class. Invalid
graph or rational encodings can first be rejected in polynomial time.

To see the counting-hierarchy containment, define
$\mathrm C_0\mathrm P=\mathrm P$,
$\mathrm C_{i+1}\mathrm P=\mathrm{PP}^{\mathrm C_i\mathrm P}$, and
$\mathrm{CH}=\bigcup_i\mathrm C_i\mathrm P$.
The standard inclusions $\mathrm{NP}^A\subseteq\mathrm{PP}^A$ and
$\mathrm P^A\subseteq\mathrm{PP}^A$ relativize, placing
$\mathrm P^{\mathrm{NP}^{\mathrm{PP}}}$ inside
$\mathrm C_3\mathrm P\subseteq\mathrm{CH}$.

\textbf{Both languages therefore belong to
$\mathrm P^{\mathrm{NP}^{\mathrm{PP}}}$, and hence to the counting hierarchy.
Theorem~\ref{thm:upper} follows.}
\end{proof}

\begin{remark}[Markov chains with rational transition probabilities]
The same proof applies to explicit irreducible rational Markov chains,
without any bound on the numerical magnitude of their expected cover times.
The hitting equations use the given rational transition probabilities;
their exact solutions still have polynomial bit length and are computable
in polynomial bit time.
\end{remark}

\section*{AI methodology}
OpenAI's ChatGPT and Codex assisted with mathematical exploration and
proof development, counterexample searches, Lean formalization,
computational checks, and manuscript preparation. Their use influenced
the constructions and arguments throughout the paper. The scope of Lean
verification and its external dependencies are described in the companion
verification document~\cite{CoverTimeLean}. For general rational Markov chains, the formalization
assumes a polynomial-time solver for the exact rational linear systems,
with a specified input encoding. The author takes responsibility for all content.
\par
\endgroup

\end{document}